# Carbon nanobuds based on carbon nanotube caps: A first-principles study

Ji Il Choi,[a]* Hyo Seok Kim,[a]* Han Seul Kim,[a] Ga In Lee,[b] Jeung Ku Kang,[a,b] and Yong-Hoon Kim[a]†

Based on density functional theory calculations, we here show that the formation of a fullerene $C_{60}$ carbon "nanobud" (CNB) on carbon nanotube (CNT) caps is energetically more favorable than that on CNT sidewalls. The dominant CNB formation mode for CNT caps is found to be the [2+2] cycloaddition reaction as in the conventional CNT sidewall case. However, it is identified to be exothermic in contrast to the endothermic reaction on CNT sidewalls. Computed reaction pathways further demonstrate that the formation (dissociation) barrier for the CNT cap-based CNB is slightly lower (significantly higher) than that of the CNT sidewall-based CNB, indicating an easier CNB formation as well as a higher structural stability. Additionally, performing matrix Green's function calculations, we study the charge transport properties of the CNB/metal electrode interfaces, and show that the $C_{60}$ bonding to the CNT cap or open end induces resonant transmissions near the Fermi level. It is also found that the good electronic linkage in the CNT cap–$C_{60}$ cycloaddition bonds results in the absence of quantum interference patterns, which contrasts the case of the CNB formed on an open-ended CNT that shows a Fano resonance profile.

## Introduction

Hybrid carbon nanostructures that integrate zero-dimensional fullerenes, one-dimensional carbon nanotubes (CNTs), and two-dimensional graphene provide an intriguing possibility to incorporate the advantageous properties of individual carbon nanoallotropes in a selective manner.[1-6] In particular, about a decade after the discovery of nanopeapod,[1] a novel fullerene-CNT hybrid nanostructure called "carbon nanobud" (CNB) has been synthesized,[2,3] in which fullerenes are covalently bonded to the outer surface of CNTs. The strong covalent coupling between fullerene and CNT as well as the location of fullerene outside of CNT could allow one to extract from CNBs functionalities that are not available for nanopeapods. Indeed, several studies demonstrated or predicted their novel electronic, magnetic, transport, and functionalization properties.[7-14]

In this work, based on density functional theory (DFT) calculations, we consider the covalent bonding of fullerene $C_{60}$ to CNT caps and show that CNT caps represent energetically much more favorable sites of CNB formation than CNT sidewalls that have been exclusively considered up to now. First, we find that the dominant CNT cap-$C_{60}$ binding motif is the [2+2] cycloaddition reaction that is similar to the one found in the $C_{60}$-$C_{60}$ binding, and the positive curvature of CNT caps induced by pentagonal defects results in the exothermic $C_{60}$ binding on CNT caps rather than the typical endothermic one found for CNT sidewalls. Furthermore, we show that the CNB formation and dissociation energy barriers for CNT caps are respectively lower and higher than those for CNT sidewalls, which implies that CNBs on CNT caps can be more easily synthesized and will be structurally more stable. In terms of computation, we will point out that more accurate estimation of energetics of CNBs it is crucial to adopt a DFT scheme designed to better describe van der Waals (vdW) interactions. In addition, we also studied the charge transport characteristics of metal-CNB interfaces. In previous theoretical works, focus were mostly made on the intrinsic properties of CNBs, and little attention was given to their interfaces with metallic electrodes. We will show that, while the CNB formed on the CNT open end shows an n-type device polarity and a Fano resonance transmission profile, the CNT cap-based CNBs exhibit strong transmission resonances near the Fermi level and no quantum interference pattern.

## Methods

All the results reported in this work were obtained based on first-principles density-functional theory (DFT) and matrix Green's function (MGF) calculations. To consider the energetics of CNT cap-$C_{60}$ binding, we have performed DFT calculations within the Perdew-Burke-Ernzerhof generalized gradient approximation (PBE GGA)[15] using the VASP package.[16] The pairwise dispersions were corrected by the DFT-D3 method by Grimme *et al.*[17] Projector augmented wave method within the plane wave basis set was used to describe the interaction between ion cores and valence electrons.[18] An energy cutoff of 400 eV for the wavefunctions and the force convergence criterion of 0.02 eV/Å were adopted. To obtain optimized geometries of capped CNTs, we used the cap models that were extended by two (10,0) CNT unit cells (80 atoms) passivated by hydrogen atoms at the open end. We adopted a tetragonal supercell with a size of 30×30×40 Å³ to avoid the unphysical interactions with periodic images. The k-point was sampled only at the Γ point within the Brillouin zone. The binding energies of CNBs were computed according to

$$E_{bind} = E(\text{C}_{60} + \text{CNT}) - E(\text{C}_{60}) - E(\text{CNT}), \quad (1)$$

where $E(\text{C}_{60}+\text{CNT})$, $E(\text{C}_{60})$, and $E(\text{CNT})$ are the total energies of geometry-optimized $\text{C}_{60}+\text{CNT}$, $\text{C}_{60}$, and CNT, respectively. Reaction energy barriers for the CNB formation on the CNT sidewall and CNT cap were calculated using a nudged elastic band (NEB) method.[19,20]

[a.] Graduate School of Energy, Environment, Water, and Sustainability, Koran Advanced Institute of Science and Technology, Daejeon 305-701, Korea.
[b.] Department of Materials Science and Engineering, Koran Advanced Institute of Science and Technology, Daejeon 305-701, Korea
\* These authors equally contributed to this work
† To whom correspondence should be addressed. E-mail: y.h.kim@kaist.ac.kr

The electronic and coherent charge transport properties of CNB-metal electrode interfaces were calculated for the symmetrically constructed $C_{60}$+CNT+$C_{60}$ junction models using our in-house MGF code[21, 22] that processes the DFT output from the SeqQuest software (Sandia National Laboratories). For the initial DFT calculations, the double-$\zeta$-polarization level basis set was adopted or the electrode Al atoms and the $C_{60}$ and CNT-cap C atoms, while the single-$\zeta$ level basis set was used for the bulk-CNT C atoms. Next, we calculated the spectral function,

$$A(E) = i[G(E) - G^+(E)], \quad (2)$$

and the transmission function,

$$T(E) = T_r[\Gamma_1(E)G(E)\Gamma_2(E)G^+(E), \quad (3)$$

where $G$ and $G$ are the retarded Green's function, $G(E) = (ES - H - \Sigma_1 - \Sigma_2)^{-1}$, and the broadening matrix, $\Gamma_{1,2} = i(\Sigma_{1,2} - \Sigma_{1,2}^+)$, respectively, and $\Sigma_{1,2}$ are the self-energy that describes the shift and broadening of the $C_{60}$+CNT+$C_{60}$ energy levels due to the coupling of $C_{60}$+CNT+$C_{60}$ with the metal electrode 1/2. Further details and accuracy check of our computational approach for the $C_{60}$-based and CNT-based junctions can be found in Refs.[23, 24] and [25-27], respectively.

## Results and discussion

### Energetics of the $C_{60}$ bonding to CNT caps

The starting point of this study is the capped (10,0) CNT (**cap**) models summarized in Fig. 1. As in half-fullerenes, carbon caps can be created by introducing six pentagons onto a hexagonal bonding network according to Euler's theorem. For the (10,0) CNT, there are seven ways to construct a cap while obeying the isolated pentagon rule [Fig. 1(a)].[28] Using these capped CNT models, we have previously shown that the metal electrode contacts based on CNT caps can achieve intrinsically low resistance (Schottky-barrier-dominated high-slope length scaling of contact resistance is absent), and it results from the uniquely good connectivity between topological defect states originating from CNT cap pentagons and CNT zigzag edge and body states.[27] In addition to the seven capped CNT models, as a reference, we considered a hydrogenated open-ended (10,0) CNT (**H-end**) model.[25, 26]

In constructing the CNT cap-based CNB models, we explored various CNT cap-$C_{60}$ bonding geometries and their energetics. All the considered bonding sites within the CNT caps are denoted by red closed circles in Fig. 1. Since one of the most favorable $C_{60}$-$C_{60}$ bonding motives is the [2+2] cycloaddition reaction of a hexagon-hexagon (66) double bond (the adjoining edges between two six-membered rings) on one $C_{60}$ to a 66 bond on the other $C_{60}$ (66/66),[23, 24, 29-35] we first considered the 66 bonding sites in the pyracylene unit arrangement (two hexagons that share a central reactive double bond are flanked by two pentagons)[36] for all the cap models except **cap 6**, for which the 66 bond is not available. Additionally, we took into account the pentagon-hexagon (56) bonding sites for all the cap models as well as several 66 bonding sites that are flanked by either one pentagon (66′; for the **cap2**, **cap3**, **cap4**, **cap5**, and **cap6** models) or no pentagon (66″; for the **cap3**, **cap4**, and **cap5** models).

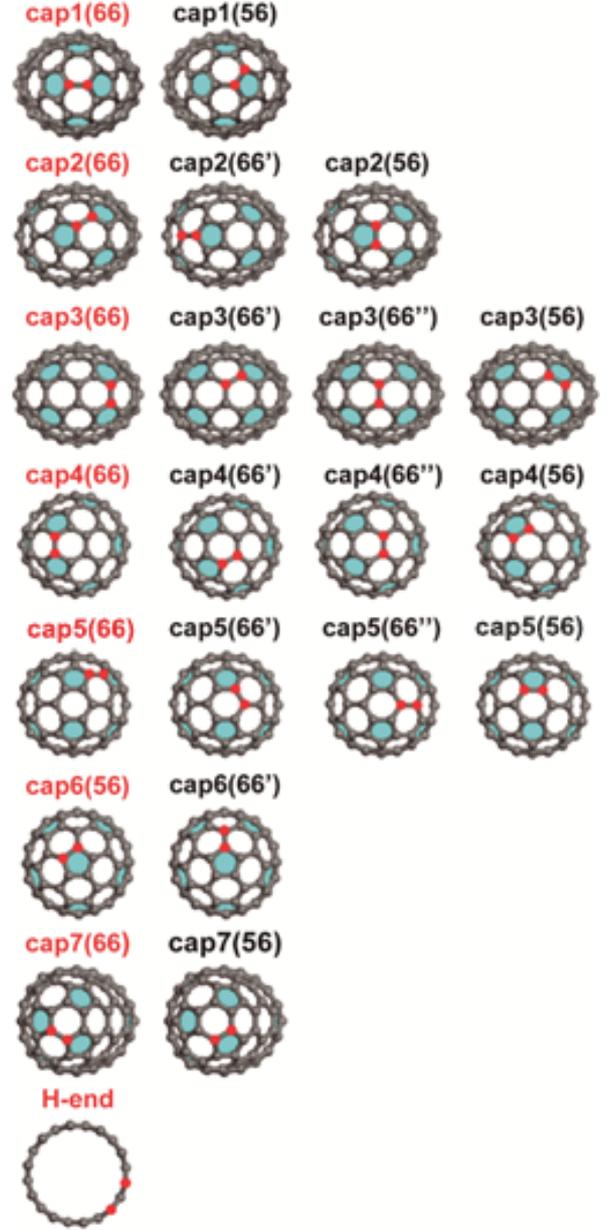

**Fig. 1.** Top views of the eight (10,0) CNT models with different end atomic structures. The red circles indicate the sites where the covalent bonding with $C_{60}$ was established to prepare CNB models based on CNT caps and open end. The pentagon topological defects within CNT caps are highlighted in cyan color. For each cap model, the leftmost one represents the energetically most favorable bonding configuration.

**Table 1.** The most preferable CNT cap-$C_{60}$ binding geometries and energies for the seven (10,0) CNT cap models shown in Fig. 1. Data for energetically less favorable CNB structures are presented in compiled in Table S1.

| Model | Cap bonding geometry | C-C bond length (Å) | Binding energy (eV) |
|---|---|---|---|
| cap1 | 66 | 1.587 | –0.276 |
| cap2 | 66 | 1.593 | –0.200 |
| cap3 | 66 | 1.590 | –0.466 |
| cap4 | 66 | 1.592 | –0.158 |
| cap5 | 66 | 1.592 | –0.224 |
| cap6 | 56 | 1.615 | –0.476 |
| cap7 | 66 | 1.588 | –0.425 |

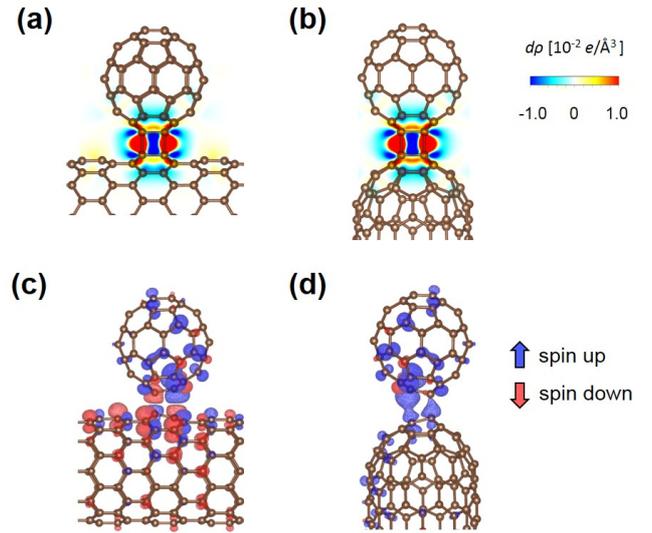

**Fig. 2.** Charge density difference plots of the CNBs based on the (a) sidewall and (b) cap (**cap4**) attachment to the (10,0) CNT. Electron accumulation (bond formation) and depletion are shown in red and blue colors, respectively. Spin densities developing at the transition states toward the formation of CNBs in the (10,0) CNT (c) sidewall and (d) **cap4** attachment cases. In (c) and (d), the magnitude of isosurface values is 0.0025 $e/Å^3$.

The calculated CNT cap-$C_{60}$ binding energies and bond lengths for the energetically most favorable configurations are summarized in Table 1 (see ESI Table S1 for the data of other binding configurations presented in Fig. 1). As expected, we find that the preferred bonding modes are the 66/66 [2+2] cycloaddition bonding except the 56/66 bonding of the **cap6** model, in which the 66/66 bonding is not available. The average bond distances between $C_{60}$ and CNT caps of the energetically favorable CNB structures are 1.59 ~ 1.61 Å. Most importantly, the chemisorption of $C_{60}$ to CNT caps is identified as an exothermic process, with the calculated binding energies ranging between – 0.158 eV and – 0.476 eV.

In comparison, the CNB formation on CNT sidewalls is an endothermic process: For the energetically favorable [2+2] cycloaddition bonding configurations, in which the bonds are established along the CNT axial directions, our calculated CNT sidewall-$C_{60}$ binding energies for the (10,0) and (5,5) CNTs were calculated to be + 0.736 eV and + 0.685 eV, respectively. We emphasize that, without pairwise vdW dispersion corrections,[17] we obtain significantly unfavorable binding trend with the binding energies of + 1.550 eV and + 1.466 eV, respectively, which are in good agreement with previous calculations.[7, 11] The underestimation of binding energies by about 0.8 eV emphasizes the necessity of including vdW effects in considering the energetics of CNBs.

To compare the cycloaddition bonding of $C_{60}$ to CNT sidewalls with that to CNT caps, we have calculated their charge density differences according to

$$\Delta\rho = \rho(C_{60}+CNT) - \rho(C_{60}) - \rho(CNT), \quad (4)$$

for the (10,0) CNT sidewall and **cap4** models. The results presented in Figs. 2(a) and 2(b), respectively, show that the local bonding natures in the CNT sidewall and CNT cap cases are quite similar and characterized by a balanced charge transfer from CNT and $C_{60}$ to the covalent bond region. For the CNBs based on other **cap** models, we obtained similar characterizations (ESI Fig. S1). We also emphasize that this charge transfer characteristic is unique to cycloaddition bonds, and significantly asymmetric charge transfer occurs in, e.g., a $C_{60}$ covalently bonded to the CNT open end (see ESI Fig. S1).

Next, we examined the process of CNB formation by calculating the minimum energy path (MEP) for the [2+2] cycloaddition reaction with the NEB approach.[19, 20] The NEB method involves optimizing a chain of images that connect the reactant (R) and product (P) states with a fictitious spring force. Each image is then allowed to move only along the direction perpendicular to the hyper-tangent, resulting in the energy being minimized in all directions except for the direction of the reaction path. Taking first the $C_{60}$ physisorbed on the sidewall of (10,0) CNTs as R and calculating the MEP toward the P configuration identified in its binding energy calculation, we obtained the formation and dissociation barriers of 2.51 eV and 1.04 eV, respectively, within PBE GGA. These are again in good agreement with the previously reported values of 2.51 eV and 0.96 eV, respectively.[7, 11] However, since we have found that the inclusion of vdW effects can significantly modify the binding energies, we repeated the MEP calculation within the DFT-D3 scheme and additionally allowing spin polarized states. The results are shown in Fig. 3. Within the spin-unpolarized DFT-D3 method, we indeed obtain much reduced formation and dissociation barrier values of 1.96 eV and 0.80 eV, respectively (see ESI Fig. S2). Inclusion of spin-polarization effects further reduces them to 1.92 eV and 0.77 eV, respectively. This indicates that, although the R and P configurations are spin-unpolarized, the transition state (TS) is spin polarized. Inspecting spin density distributions around TS, we find that ferromagnetic spin states develop on the bonding regions of $C_{60}$ and CNT sidewalls and the two spin-polarized states are antiferromagnetically aligned [Fig. 2(c)].

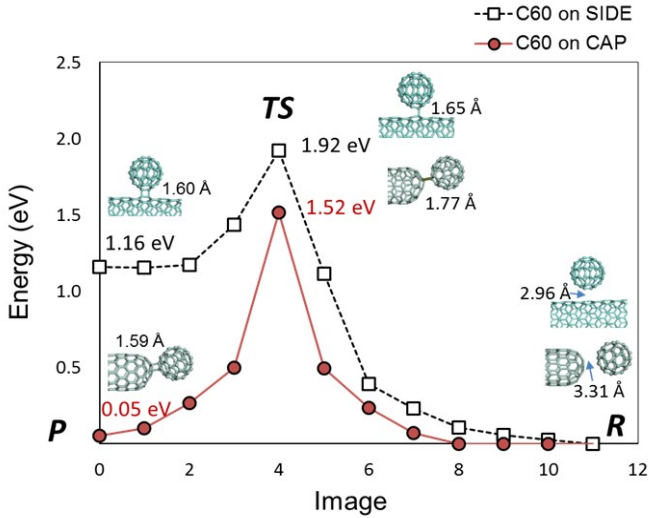

**Fig. 3.** Minimum-energy path for the formation of C$_{60}$-**cap4** (66) configuration. The *R*, *TS*, and *P* symbols denote the reactant, transition states, and product, respectively. The energy of the reactant was set to zero.

**Table 2.** Binding energies, formation energy barriers, and dissociation energy barriers for the C$_{60}$ reactions on the sidewall and cap regions of (10,0) CNT calculated with and without DFT-D3 dispersion corrections.

| Model | Binding energy (eV) | | Formation barrier (eV) | | Dissociation barrier (eV) | |
|---|---|---|---|---|---|---|
| | DFT-D3 | PBE | DFT-D3 | PBE | DFT-D3 | PBE |
| CNT sidewall | 0.736 | 1.550 | 1.92 | 2.51 | 0.77 | 1.04 |
| CNT sidewall[7] | - | 1.558 | - | 2.51 | - | 0.96 |
| CNT cap (**cap4**) | −0.158 | 0.436 | 1.52 | - | 1.46 | - |

We now discuss the MEP for the CNB formation on CNT caps by taking the **cap4** model as the representative case. We adopted the configuration in which C$_{60}$ is physisorbed on the CNT **cap4** as the *R* structure and that presented earlier in Fig. 1 as the *P* structure. Performing spin-polarized DFT-D3 calculations (see ESI Fig. S2 for the spin-unpolarized calculation result), we obtained the formation energy barrier of 1.52 eV and the dissociation energy barrier of 1.46 eV (Fig. 3 and Table 2). Or, compared with the CNT sidewall case, the CNB formed on the CNT cap has the formation energy barrier that is lower by 0.46 eV and the dissociation barrier that is higher by 0.69 eV. So, we can conclude that CNT caps represent more preferable and yet sustainable CNB formation sites than CNT sidewalls.

A few comments are in order. First, we emphasize that, rather than the simple curvature effect, the presence of pentagonal defects endows CNT caps with the favorable CNB formation energetics. To create carbon caps, six pentagonal rings must be introduced according to Euler's theorem. These topological defects generate localized states near the Fermi level,[37-39] and they can be expected to facilitate the formation CNBs. This can be indirectly seen in the energetic ordering of different CNB configurations (Fig. 1 and ESI Table S1), which shows the general preference for the binding of C$_{60}$ near the pentagonal sites on CNT caps. Thus, although it might become more difficult to form CNBs as the CNT diameter increases,[7, 11] we expect that the CNB formation on CNT caps would be generally preferable over that on CNT sidewalls for a given CNT diameter.

Next, observing the spin polarization at the *TS* state, we again find that ferromagnetic states develop at the C$_{60}$ and CNT cap interface regions. However, unlike the CNT sidewall counterpart, we interestingly find that the two are ferromagnetically aligned in the CNT cap case [Fig. 2(d)]. Noting that magnetic states were theoretically predicted to be stabilized in CNBs fused on CNT sidewalls,[10, 13] we expect that magnetic CNBs can be also synthesized by fusing CNBs on CNT caps. The difference in their *TS* might also indicate that unique magnetic states are available for CNBs formed on CNT caps, but we leave these considerations for future work.

### Electronic and charge injection properties of the metal/C$_{60}$+CNT interfaces

CNBs have been suggested to be useful for various applications such as field-emission display[2] and Li ion battery[40], and it was recently announced that flexible touch sensors based on transparent films containing CNBs are near commercialization[41]. We here study the charge transport across metal-CNB interfaces, which could be related with cold electron field-emission and STS measurements[2] and provide useful information for future CNB-based device applications. We particularly compare the CNBs based on CNT caps and open-ended CNT to reveal the uniqueness of CNBs formed on CNT caps.

In. Fig. 4, we show the atomic structures and corresponding electronic structure of symmetric CNB models (C$_{60}$+CNT+C$_{60}$) based on the **cap3** and **H-end** cases in contact with two metal electrodes modeled by the 6×6 Al(111) slabs. To avoid any crosstalk between the two metal-C$_{60}$ interfaces,[26, 27] we have used sufficiently long junction models with the channel lengths of 6.01 nm and 6.26 nm (CNT lengths of 3.99 nm and 4.33 nm) for the CNBs based on **cap3** and **H-end**, respectively (ESI Table S2). In the local density of states (DOS) plots of the **cap3** and **H-end** models, we separately focus on the metal-C$_{60}$ and C$_{60}$-CNT interfaces. In terms of the metal-C$_{60}$ interfaces, we observe the pining of the C$_{60}$ lowest unoccupied orbitals at right above the Fermi level $E_F$, significant hybridization between C$_{60}$ and Al states, and formation of large metal-induced gap states. These result from the highly electron-accepting nature of C$_{60}$s and the corresponding large amount of charge transfer from metals to CNBs.[23, 24] We have indeed confirmed that these features are consistently observed for all other **cap** CNB models (ESI Fig. S3), and that they do not modify the nature of covalently bonded C$_{60}$-CNT interfaces (ESI Fig. S4).

While the C$_{60}$ local DOS in the **cap3** and **H-end** CNB models can be similarly characterized, we find notable differences in their CNT local DOS. This indicates that the nature of C$_{60}$-CNT interfaces are different in the two cases. More specifically, as shown in ESI Figs. S1 and S4, a large amount of electron transfer from CNT to C$_{60}$ in the **H-end** model results in the downshift of CNT states and corresponding $n$-type polarity of the combined C$_{60}$+CNT system (Fig. 4b). On the other hand, a balanced and much smaller charge redistribution occurring in the **cap3** CNB model (ESI Figs. S1 and S4) induces a bipolar character (Fig. 4a). Finally, in Fig. 4, we observe near $E_F$ strong DOS that propagate (Fig. 4a) or decay (Fig. 4b) into the CNT body (denoted by red left triangles). These states originate from the CNT cap [27] and open CNT end [26] states in the **cap3** and **H-end** CNB models, respectively.

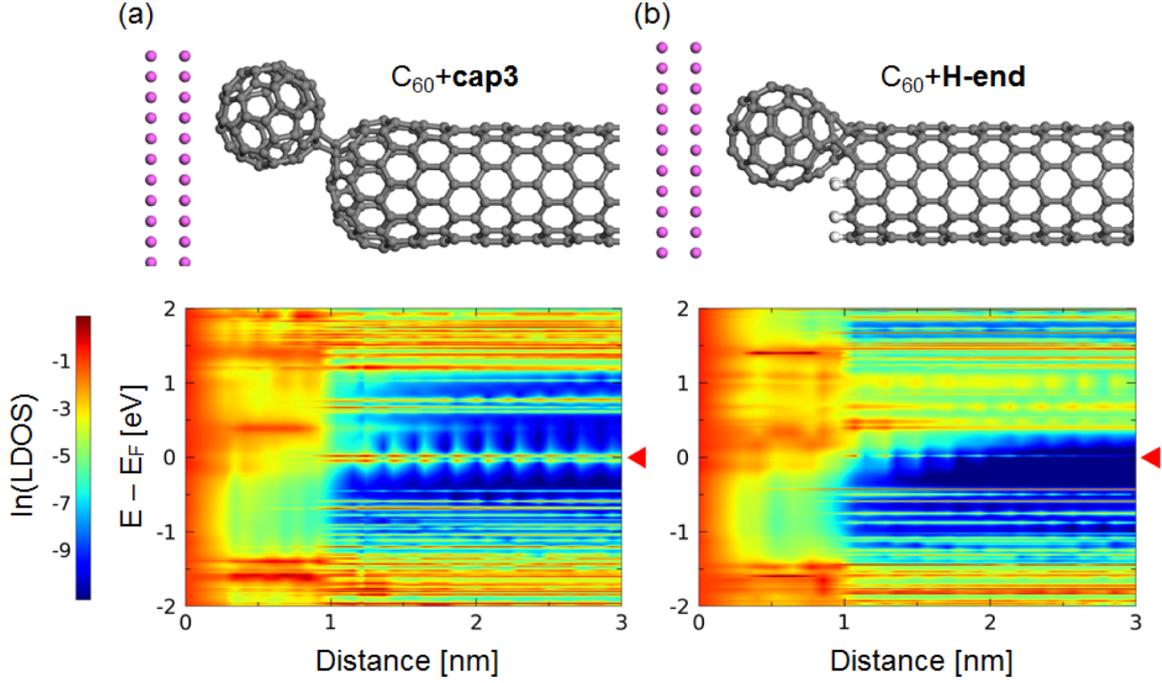

**Fig. 4.** Side views (top panels) of the (a) Al(111)-C$_{60}$+**cap3** and (b) Al(111)-C$_{60}$+**H-end** interfaces and their local DOS along the channels (bottom panels). The red left triangle in (a) indicates the CNT cap-originated topological defect states that propagate into the CNT body, while that in (b) indicates the localized CNT edge states that exponentially decay into the CNT body.

We now consider how the spatial and energetic distributions of CNB electronic states affect the interface charge transport properties. In Fig. 5, we present the transmission spectra obtained from the two junction models, together with the DOS projected on two C$_{60}$s (orange color), two CNT caps or hydrogenated CNT ends (blue color), and the CNT body (green color). In line with the above discussion, we observe the overall bipolar and $n$-type transport characters of the **cap3** and **H-end** junction models. In addition, as the most notable features, sharp resonant transmission peaks appear near $E_F$, which originate from the CNT cap and CNT end states in the **cap3** and **H-end**-based CNB models, respectively (denoted by red left triangles in Fig. 4). Focusing on the details, we note that the propagating CNT cap states (well-connected with CNT body states) in the former[27] induce much larger transmissions than the exponentially decaying hydrogenated CNT end states in the latter.[26] In spite of the differences, comparing with the data obtained from CNTs with open-end [26] and cap structures,[27] we can generally conclude that the fullerene attachment at CNT end regions can induce energetically (and spatially) localized strong electron transmissions near $E_F$.

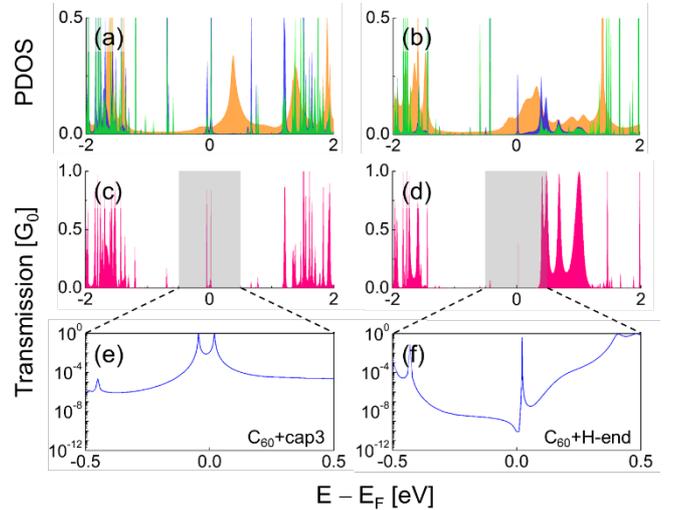

**Fig. 5.** Projected DOS of the (a) C$_{60}$+**cap3**+C$_{60}$, and (b) C$_{60}$+**H-end**+C$_{60}$ junction models. Orange, blue, and green colors represent the projections onto two C$_{60}$s, two CNT caps (C$_{60}$+**cap3**+C$_{60}$) or hydrogenated CNT one unit-cell ends (C$_{60}$+**H-end**+C$_{60}$), and the CNT body, respectively. Corresponding transmission functions are shown in (c) and (d), respectively. In the bottom panels (e) and (f), transmissions near the Fermi level are reproduced in a logarithmic scale.

We finally address an interesting question on the presence of quantum interference features in transmission spectra. The metal-$C_{60}$ and $C_{60}$–CNT interfaces were suggested to act as double barriers and result in constructive and destructive interference transmission patterns.[42]. To check the above hypothesis, we observed the transmission spectra near $E_F$ in a logarithmic scale as shown in Figs. 5(e) and 5(f). We find that, while Fano resonance features[43] exist in the CNB model based on **H-end**, CNBs based on capped CNT models do not produce any interference pattern. The asymmetric line profile in the **H-end**–based CNB model, the main characteristic of the Fano resonance, results from the destructive interference between the modes directly propagating from CNT to electrodes and the states localized at $C_{60}$. On the other hand, its absence in the junction models based on capped CNTs implies that the double-barrier model is not applicable, which reconfirms the good electronic connectivity between cycloaddition-bonded CNT caps and $C_{60}$s.

## Conclusions

In summary, carrying out first-principles calculations, we have established that the $C_{60}$ [2+2] cycloaddition functionalization of CNT caps is significantly more favorable over that of CNT sidewalls or graphene basal plane. This result should have important practical implications in that the high energetic cost of CNB formation and associated complexity of CNB synthesis [2,3] have been so far a major challenge for widespread CNB applications. In terms of computational aspects, we confirmed that weak dispersion interactions should be properly taken into account to accurately assess the energetics of hybrid carbon nanostructures. An increased availability of CNBs should expedite their device applications, and in this context we next focused on their charge injection/emission properties across the CNB-metal interfaces. We found that the CNB formation at CNT ends can generally induce strong and localized electron transmissions near the Fermi level. We further pointed out that the good electronic connectivity of the CNT cap-$C_{60}$ manifests as the absence of Fano resonance profiles in the transmission spectra.

## Acknowledgements

This work was supported mainly by the Global Frontier Hybrid Interface Materials Program (2013M3A6B1078881) of the National Research Foundation (NRF) funded by the Ministry of Science, ICT and Future Planning of Korea, and additionally by the NRF Nano·Material Technology Development Program (2012M3A7B4049888) and the KIST Institutional Program (2Z04490-15-112). Computational resources were provided by the KISTI Supercomputing Center (KSC-2014-C3-021).